\documentclass[aps,pre,twocolumn,showpacs,superscriptaddress,amssymb,amsmath]{revtex4}
\usepackage{graphicx}
\newcommand{\be}{\begin{equation}}
\newcommand{\ee}{\end{equation}}

\begin{document}

\title{\bf Heat generation and transport due to time-dependent forces}
\author{Bijay Kumar Agarwalla} 
\affiliation{Department of Physics and Centre for Computational Science and Engineering, National University of Singapore, Singapore, 117542, Republic of Singapore}
\author{Jian-Sheng Wang} 
\affiliation{Department of Physics and Centre for Computational Science and Engineering, National University of Singapore, Singapore, 117542, Republic of Singapore}
\author{Baowen Li}  
\affiliation{Department of Physics and Centre for Computational Science and Engineering, National University of Singapore, Singapore, 117542, Republic of Singapore}
\affiliation{NUS Graduate School for Integrative Sciences and Engineering, Singapore 117456, Republic of Singapore}

\date{\today}

\begin{abstract} 
  We study heat generation and transport properties for solids in the presence of arbitrary time-dependent force. Using nonequilibrium Green's function (NEGF) approach we present an exact analytical expression of heat current for the linear system. We found that the current can be expressed in terms of the displacement of the atoms in the center and the self energy of the heat bath. We carry out the calculation for periodic driven force and study the dependence of steady state current with frequency and system size for one and two-dimensional systems. We obtain an explicit solution of current for one-dimensional linear chain connected with Rubin bath. We found that the heat current is related to the density of states of the system and is independent of the bath temperature in ballistic transport. The baths can absorb energy only when the external frequency lies within the phonon band frequency. We also discuss the effect due to nonlinear interactions in the center.
\end{abstract}


\maketitle

\section{Introduction}
In recent years, the understanding of heat transport in mesoscopic systems has drawn a lot of attention because of their interesting physical properties and vast applications ranging from nanosize electronic devices to thermal transistors. In addition there has been a great deal of interest to study how to manipulate and control heat. Different theoretical models have been proposed to control thermal transport \cite{casati,baowen1,baowen2}. Several experimental works have also been carried out \cite{chang,xie}. To understand the generic features of these systems numerous studies have been done using nonequilibrium Green's functions (NEGF) \cite{wang06,wang08,eduardo,eduardo1}, generalized Langevin equation \cite{Dhar,Dhar1} and quantum master equation \cite{Tanimura,gaspard} approach.

The energy transport in general can be achieved as a response to the temperature gradient or chemical potential gradients and in the linear response regime is governed by Fourier's law \cite{Dhar, Lepri, Lebowitz} for diffusive systems. It is also expected that time-dependent external force can induce directed heat transport between the leads at the same temperature or even in the presence of temperature gradient \cite{Ren,Li1,Li2}. However, whether all energy driven by external force can be transmitted to the reservoir or not is a valid question. Recent study on driven quantum Langevin model for any arbitrary time-dependent potential shows that the energy dissipation flow to thermal environment is related to the violation of the fluctuation-response relation \cite{saito, Harada}. Understanding the general features of current is therefore one of the main goals in nonequilibrium statistical physics.  

For systems driven arbitrarily far from equilibrium it is possible to relate the work done during the nonequilibrium process with the free energy difference between two equilibrium states through Jarzynski's equality (JE) \cite {JE, JE1} which states that 
\begin{equation}
\langle e^{-\beta W}\rangle=e^{-\beta \Delta F},
\label{eq:JE}
\end{equation}
where $W$ is the work done (here the work done is due to external time-dependent force) and $\Delta F$ is the difference of free energy between final and initial equilibrium states. The average is over the work distribution function $P(W)$ and $\beta=1/(k_{B}T)$. If $P(W)$ is Gaussian then it can be shown for classical systems that $\langle W \rangle=\Delta F + \beta \sigma_{W}^{2} /2 $ where $\sigma_{W}^{2}=\langle W^{2}\rangle-\langle W \rangle^{2}$ is the variance. An important point to realize in this case is that even if the average work $\langle W \rangle$ is independent of temperature the variance increases linearly with temperature. Finding explicit forms of the nonequilibrium distribution functions and henceforth averages for different systems is of obvious interest to verify JE \cite{fluc2, fluc3, fluc4}.

In this paper we investigate the influence of the external time-dependent force on a harmonic system which is connected with heat baths and analyze the energy current with the applied frequency and system size. We explore the effect on current due to two different types of heat baths, Rubin \cite{Rubin} and Ohmic \cite{Weiss}. We discuss briefly that one-dimensional linear chain model can not be used as a heat pump. To obtain the expression for current and to examine the underlying physical process we use NEGF method. Aiming for an exact analytical solution of current we consider special form of time-dependent potential which is linear in system's position coordinates.

The paper is organized as follows. In the next section, we introduce our model and derive the expression of energy current in time domain which in general is true for any form of time-dependent force and also in any dimension. In Sec. III we choose periodic driven force and study steady state properties for one-dimensional (1D) linear chain and two-dimensional (2D) square lattice. We present an explicit solution for current in one-dimensional linear chain which is connected to Rubin baths. In Sec. IV we discuss the effect on heat current due to nonlinear interaction in the center. Finally we conclude with a short discussion in Sec. V.

\section{The Model}
We consider an insulating solid where only the vibrational degrees of freedom
plays important role for heat transport. Our model consists of a finite harmonic center which we denote by $C$, 
coupled to two heat baths ($L$ and $R$) kept at temperatures $T_L$ and $T_R$. For the heat baths we consider the standard model of an infinite 
collection of oscillators. Let the displacement
from some equilibrium position for the $j$-th degree of freedom in the
region $\alpha$ be $u_j^\alpha$, $\alpha = L, C, R$. The 
Hamiltonian is given by
\begin{equation}
{\cal H}={\cal H}_L+{\cal H}_C+{\cal H}_R+ {\cal H}_{LC}+{\cal H}_{RC}+{\cal V}(t),
\end{equation}
where
\begin{eqnarray}
&&{\cal H}_{\alpha}= \frac{1}{2}
{(\dot{u}^\alpha)}^T \dot{u}^\alpha + \frac{1}{2} {(u^\alpha)}^T
K^\alpha u^\alpha,  ~~~~~~~~\alpha=L,C,R  \nonumber\\
&&{\cal H}_{\alpha C}=(u^\alpha)^T V^{\alpha C} u^C,  ~~~~~~~~~~~~~~~~~~~~~~~~~~\alpha=L,R \nonumber \\
\end{eqnarray} 
where superscript $T$ denotes matrix transpose, $u^\alpha$ is a column vector consisting of all
the displacement variables in region $\alpha$, and $\dot{u}^\alpha$ is
the corresponding conjugate momentum.  $K^\alpha$ is the spring
constant matrix and $V^{\alpha C}=(V^{C \alpha})^T$ $(\alpha=L,R)$ is the coupling matrix of the
leads to the central region. ${\cal V}(t)$ is the time-dependent external potential which depends only on the center atom variables. In this case the potential has a particular form $-\theta(t-t_{0})f^{T}(t)u^{C}$ and $f(t)$ is the time dependent force vector acting only on center atoms. The force can be in the form of an applied electromagnetic field. For simplicity we have set all the atomic masses to 1, but the formulas can be used for variable masses with a transformation $u_{j}\rightarrow x_{j}\sqrt{m_j}$. We assume that at $t<t_{0}$ the system is under a known nonequilibrium steady state $\rho$ with respect to the Hamiltonian ${\cal H}_{0}$ where ${\cal H}_{0}$ is the Hamiltonian without the time-dependent potential ${\cal V}(t)$. For $t \geq t_{0}$ the time-dependent force drives the system into a nonequilibrium state. We are interested in calculating the current going from the left lead to the center.

The energy current flowing out of the left lead is given by
\be
I_{L}(t) = -\left< \frac{d{\cal H}_{L}(t)}{dt} \right> = \frac{i}{\hbar}
  \left< \left[ {\cal H}_{L}(t),{\cal H}(t) \right] \right>,
\ee
where the average is with respect to the density operator $\rho$ defined above. The operators are in Heisenberg picture. The position and momentum operators obey the canonical commutation relation
\begin{equation}
\left[ u_j^{\alpha}(t), \dot{u}_k^{\beta}(t) \right] = i \hbar\,\delta_{jk}
  \,\delta^{{\alpha}{\beta}},~~~~~\alpha, \beta = {L, C, R}.
\label{eq:commutation}
\end{equation}
\begin{figure}
\includegraphics[width=0.5\columnwidth]{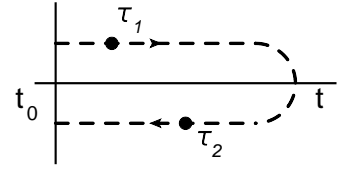}%
\caption{\label{fig3}The complex-time contour in the Keldysh formalism. The path of the contour begins at time $t_{0}$, goes to time $t$, and then goes back to time $t=t_{0}$. $\tau_{1}$ and $\tau_{2}$ are complex-time variables along the contour.} 
\end{figure}
Equation(4) therefore becomes
\be
I_{L}(t)=\left<\dot{u}^{T}_{L}(t) V^{LC} u_{C}(t) \right>=i \hbar\, {\rm{Tr}}\left [\frac{\partial}{\partial t'}G^{<}_{LC}(t',t) V^{CL}\right]_{t'=t}
\ee
Since $[\dot{u}_{L}(t),u_{C}(t)]=0$, the above equation can also be written as 
\be
I_{L}=i\hbar\, {\rm {Tr}} \left[\frac{\partial}{\partial t'} G^{>}_{LC}(t',t) V^{CL}\right]_{t'=t},
\ee
which after symmetrization finally reduces to
\be
I_{L}=i\hbar\, {\rm {Tr}} \left[\frac{\partial}{\partial t'} \bar{G}_{LC}(t',t) V^{CL}\right]_{t'=t},
\label{maineq}
\ee
where $\bar{G}(t,t')=\frac{1}{2}\left[G^{<}(t,t')+G^{>}(t,t')\right]$. The lesser ($G^{<}$) and greater ($G^{>}$) Green's functions are defined as
\begin{eqnarray}
G_{jk}^{{LC},<}(t,t')=-\frac{i}{\hbar}\left<u_{k}^{C}(t') u_{j}^{L}(t) \right> \nonumber \\
G_{jk}^{{LC},>}(t,t')=-\frac{i}{\hbar}\left<u_{j}^{L}(t) u_{k}^{C}(t') \right>.
\end{eqnarray}

\noindent
Equation~(\ref{maineq}) is the primary equation we use to calculate for current flowing out of the left lead. To compute the current we need to determine $G_{LC}$. Our main task would be to eliminate the reference to the lead Green's functions in terms of the Green's functions of the central region. We use contour-ordered Green's function, defined on a Keldysh contour \cite{wang08,schwinger-keldysh,rammer86,jauho94} (see Fig.~1) from $t_{0}$ to $t$ and back. The contour ordered Green's function can be mapped onto four different normal time
Green's functions by $G^{\sigma\sigma'}(t,t') = \lim_{\epsilon \to
0^+} G(t\! +\! i \epsilon \sigma, t'\!+\! i\epsilon \sigma')$, where
$\sigma = \pm (1)$, and $G^{++} = G^{t}$ is the time ordered Green's
function, $G^{--} = G^{\bar{t}}$ is the anti-time ordered Green's
function, $G^{+-} = G^{<}$, and $G^{-+} = G^{>}$.  The retarded
Green's function is given by $G^r = G^t - G^{<}$, and the advanced by
$G^a = G^{<} - G^{\bar{t}}$.  These relations also hold for the self
energy discussed below. It can be shown from the equations of motion that the contour ordered Green's function for this model satisfies the equation
$G_{CL}(\tau, \tau') =\int d\tau'' G_{CC}(\tau, \tau'') V^{CL} g_L(\tau'', \tau')$, where
the integral is along the contour.  The function $g_L$ is the contour
ordered Green's function for the semi-infinite free left lead in
equilibrium at temperature $T_L$. Using Langreth's theorem \cite{jauho94} in Eq.~(\ref{maineq}) we can get
\begin{eqnarray}
I_L(t)&=& i \hbar\, {\rm {Tr}}\biggl[ \int_{t_{0}}^{\infty} dt'' \frac{\partial }{\partial t'}\Bigl[G_{CC}^{r}(t,t'')\bar{\Sigma}_{L}(t''-t')\nonumber \\
&+& {\bar{G}}_{CC}(t,t'')\Sigma_{L}^{a}(t''-t')\Bigr]_{t'=t}\biggr],
\label{current1}
\end{eqnarray}
with $\Sigma_{L}=V^{CL} g_L V^{LC}$ being the self energy due to the interaction with the left lead. The important point to note is that $G_{CC}$ does not have time-translational invariance because of the presence of time-dependent force whereas the surface Green's function $g_L$obeys this invariance as it is calculated at equilibrium. Our main task now is to calculate the center Green's function.

Let us first consider the one-point contour-ordered Green's function for the center which is defined as \cite{wang08}
\be 
G_{j}^{C}(\tau)=-\frac{i}{\hbar}\langle T_{c} u_{j}^{C}(\tau) \rangle,
\ee
where $T_{c}$ is the contour-ordering operator. For one-point Green's function contour ordering is not important. $u_{i}^{C}(\tau)$ is the operator in the Heisenberg picture. Transforming to the interaction picture with respect to the Hamiltonian ${\cal H}_{0}$ and taking the interaction Hamiltonian as ${\cal V}(t)=-\theta(t-t_{0})f^{T}(t)u^{C}$ we can write the contour ordered Green's function as 
\be
G_{j}^{C}(\tau)=-\frac{i}{\hbar}\left<T_{c} u_{j}^{C}(\tau)e^{\sum_{k}\frac{i}{\hbar} \int d\tau' f_{k}(\tau') u_{k}(\tau')} \right>_{G_{0}},
\ee
where $G_{0}$ is the Green's function calculated with the Hamiltonian ${\cal H}_{0}$. Now if we expand the exponential function, the terms with odd numbers of $u^{C}$ will be zero since the average is with respect to a quadratic Hamiltonian. So the expression will contain terms with even number of $u^{C}(\tau)$ and odd number of $f(\tau)$ and finally can be written in the matrix form as 
\be
G^{C}(\tau)=\frac{i}{\hbar}\int d\tau' G_{0}(\tau,\tau') f(\tau') + \mathrm{higher\, order\, terms},
\ee
(For notational simplicity we have omitted the superscript $CC$ on the two-point Green's function of center). In Fig.~2 we draw Feynman diagrams for $G_{i}^{C}(\tau)$ upto third order of force. The contribution from the first diagram is nonzero. However, all the higher order terms contain the same type of vacuum diagrams which are zero. Vacuum diagram in this case is defined as a diagram where all variables are integrated and the result is independent of space or time. The expression for such a diagram in terms of contour variable can be written as
\begin{eqnarray}
&&\int \int d\tau d\tau'f^{T}(\tau) G_{0}^{CC}(\tau,\tau') f(\tau') \nonumber \\
&&=\sum_{\sigma,\sigma'} \int \int \sigma dt \, \sigma' dt'{f^{\sigma}(t)}^T G_{0}^{\sigma,\sigma'}(t,t') f^{\sigma'}(t').
\end{eqnarray}
 The last line is obtained by going to the real time using Langreth's rule. Since the driven force $f$ does not depend on the branch index $f^{+}(t)=f^{-}(t)=f(t)$, we can take the summation inside and obtain \cite{wang08}
\be
\sum_{\sigma,\sigma'} \sigma \sigma' G_{0}^{\sigma,\sigma'}= G_{0}^{t}+G_{0}^{\bar{t}}-G_{0}^{<}-G_{0}^{>}=0.
\ee

So the above expression is zero. Similarly all the higher order diagrams doesn't contribute to the one-point Green's function. So the exact expression for $\left<u_{C}(\tau)\right>$ is now given by
\begin{figure}
\includegraphics[width=\columnwidth]{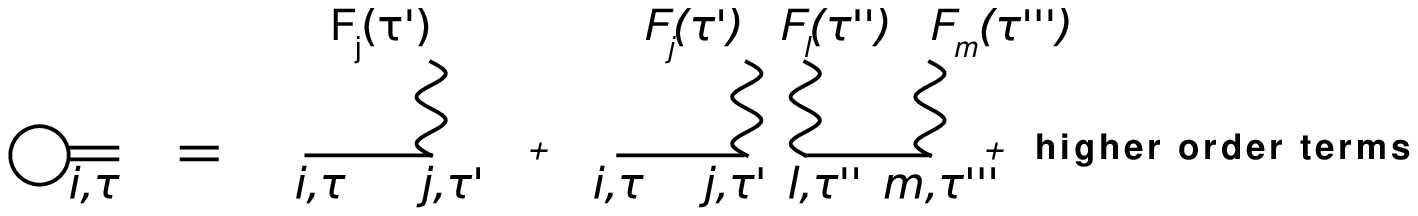}%
\caption{\label{fig3} The Feynman diagram for one-point Green's function of the center in the presence of time-dependent force.}
\end{figure}
\be
\langle u_{C}(\tau)\rangle=-\int d\tau' G_{0}(\tau,\tau') f(\tau').
\ee
From this expression it is also clear that $\left<u_{C}(\tau)\right>$ does not depend on the branch index i.e. $\left<u_{C}^{+}(t)\right>=\left<u_{C}^{-}(t)\right>$. So in real time we obtain 
\be
\langle u_{C}(t)\rangle=-\int dt' G_{0}^{r}(t-t') f(t').
\label{uc}
\ee
where $G_{0}^{r}$ is the retarded Green's function and is defined as
\be
G^{r}_{0,jk}(t,t')=-\frac{i}{\hbar} \theta(t-t')  \bigl <[u_{j}^{C}(t),u_{k}^{C}(t')] \bigr >.
\ee
It is also related to the response function in the linear response theory. In fact, the same result, Eq.~(17), can also be derived from the standard linear response theory. Similarly the two-point Green's function in the interaction picture is also calculated using the definition and is given by
\be
G_{jk}(\tau,\tau')=-\frac{i}{\hbar}\langle T_{c} u_{j}^{C}(\tau) u_{k}^{C}(\tau')e^{\sum_{m}\frac{i}{\hbar} \int d\tau'' f_{m}(\tau'') u_{m}(\tau'')} \rangle_{G_{0}}.
\ee

As discussed above we can expand the exponential and the terms greater then $\mathcal{O}(f^{2})$ vanishes as they contain same type of vacuum diagrams. The exact expression can be written as
\begin{eqnarray}
G_{jk}(\tau, \tau')&=&G_{0,jk}(\tau, \tau')-\frac{i}{\hbar}\sum_{ms}\int d\tau_{1} d\tau_{2} G_{0,jm}(\tau, \tau_{1})\nonumber \\
&&G_{0,ks}(\tau', \tau_{2})f_{m}(\tau_{1}) f_{s}(\tau_{2}).
\end{eqnarray}
In terms of $\left<u_{C}(\tau)\right>$ the center Green's function now become

\be
G(\tau, \tau')=G_{0}(\tau, \tau')-\frac{i}{\hbar} \langle u_{C}(\tau)\rangle \langle u_{C}(\tau')\rangle^{T}.
\label{green1}
\ee
From the above equation we can write  $G=G_{0}+\delta G$  with $\delta G=-\frac{i}{\hbar} \langle u_{C}(\tau)\rangle \langle u_{C}(\tau')\rangle^{T}$. Now using the property of $\langle u_{C}(\tau) \rangle $ we can write 
\be
\delta G^{++}=\delta G^{+-}=\delta G^{-+}=\delta G^{--}
\ee
which implies that $\delta G^{r}=\delta G^{a}=0$ and $\delta G^{<}=\delta G^{>}=\delta \bar{G}=-\frac{i}{\hbar} \langle u_{C}(t)\rangle \langle u_{C}(t')\rangle^{T}$.
So using Eq.~(10) the expression for the current reduces to
\begin{eqnarray}
I_L(t)&=& i \hbar{\rm{Tr}}\biggl[ \frac{\partial}{\partial t'}\int_{t_{0}}^{t} dt''\Bigl[\bar{\Sigma}_{L}(t'-t'')G_{0}^a(t'',t) \nonumber \\
&+& \Sigma_{L}^{r}(t'-t'')\bar{G_{0}}(t'',t)\Bigr]_{t'=t}\biggr]\nonumber \\
&+&{\rm{Tr}}\left[\int_{t_{0}}^{t} dt''\langle u_{C}(t)\rangle \langle u_{C}(t'')\rangle^{T}\frac{\partial }{\partial t'}\Sigma_{L}^{a}(t''-t')\right]_{t'=t} \\
&=& I_{L}^{s}(t)+I_{L}^{d}(t). \nonumber 
\label{final}
\end{eqnarray}
By writing $I_L(t)$ in this form it is clear that the contribution to the energy current is separated into two parts. $I_{L}^{d}(t)$ is the current due to  driven force and $I_{L}^{s}(t)$ is due to the temperature difference between the heat baths. In the long time limit i.e, $t \rightarrow \infty$, $I_{L}^{s}(t)$ is the steady-state heat flux and is given by the Landauer like formula \cite{wang08}
\be
I_{L}^{s}=\frac{1}{4 \pi} \int_{-\infty}^{\infty} d\omega \, \hbar\,\omega \,{\rm T}[\omega]\,(f_L-f_R)
\ee
where $f_{\alpha}=1/(e^{\beta_{\alpha} \hbar \omega}-1)$ is the Bose-Einstein distribution function where $\beta_{\alpha}=1/(k_{B}T_{\alpha})$, ${\rm T}[\omega]$ is known as the transmission function and is given by the Caroli formula $ {\rm T}[\omega]={\rm Tr}[G_{0}^{r}\Gamma_{L}G_{0}^{a}\Gamma_{R}]$ with $\Gamma_{\alpha}=i (\Sigma_{\alpha}^{r}-\Sigma_{\alpha}^{a})$ and $G_{0}^{a}[\omega]=\bigl[G_{0}^{r}[\omega]\bigr]^{\dagger}$, $\alpha=L,R$. The separation of energy current into two parts is possible because the system is linear and the driving force is not correlated with the heat baths.

If we take the two heat baths to be at the same temperature, i.e, $\Delta T=T_L-T_R=0$, then $I_{L}^{s}$ is zero. 
So in the linear case the final expression for current with $\Delta T=0$ is
\be
I_L^{d}(t)={\rm{Tr}}\left[ \int_{t_{0}}^{t} dt''\langle u_{C}(t)\rangle \langle u_{C}(t'')\rangle^{T}\frac{\partial }{\partial t'}\Sigma^{a}_{L}(t'',t')\right]_{t'=t}
\label{final1}
\ee
where $\Sigma_{L}^{a}(t''-t')=0$ if $t''-t'\geq 0$.
This is the central equation which can be used to calculate the current both in transient state as well as in steady state with arbitrary form of force. This expression is true for systems with finite heat baths and also in higher dimensions.
 
In the following we will consider situation for $\Delta T=0$ and use Eq.~(\ref{final1}) to calculate the current. We take a particular form of force which is oscillatory and carry out calculation for 1D linear chain and 2D square lattice for two types of heat baths (1) Rubin bath and (2) Ohmic bath.

\section{Periodic driven force}
We consider the form of force given by $f(t)=f_{0}e^{-i\Omega t}+c.c$ where $f_{0}$ is a column vector with complex amplitude and $\Omega$ is the driven frequency. Then from Eq.~(\ref{uc}) $\left<u_{C}(t)\right>$ can be written as
\be
\left<u_{C}(t)\right>=G_{0}^{r}[\Omega] f_{0} e^{-i\Omega t} + c.c .
\ee
where $G_{0}^{r}[\Omega]$ is given by
\be
G_{0}^{r}[\Omega]=\bigl[(\Omega+i\eta)^{2} I -K^{C}-\Sigma_{L}[\Omega]-\Sigma_{R}[\Omega]\bigr]^{-1}
\ee
with $\eta \rightarrow 0^{+}$ and $I$ is the identity matrix. We set $t_{0}\rightarrow -\infty$ for steady state oscillation and finally average over a time period $\bar{I}_L= \frac{1}{\tau} \int_{0}^{\tau}I_L(t)\,dt $ where $\tau=2\pi/\Omega$ is the time period of the driving field, we finally get from Eq.~(\ref{final1}),
\begin{eqnarray}
\bar{I}_L&=&-\Omega \, S[\Omega], \\
S[\Omega]&=&{\rm Tr}(G_{0}^{r}[\Omega]f_{0}f_{0}^{\dagger}G_{0}^{a}[\Omega]\Gamma_{L}[\Omega]),
\end{eqnarray}
Since $\Omega S[\Omega]$ is always positive the current is flowing into the lead. The average rate of work done is positive and consistent with the second law of thermodynamics. For this particular case the same result can also be obtained using linear response theory.  We can write Eq.~(28) in another form by using the following relation between $G_{0}^{r}$ and $G_{0}^{a}$
\be
G_{0}^{r}[\Omega]-G_{0}^{a}[\Omega]=-i\, G_{0}^{r}[\Omega]\,\bigl(\Gamma_{L}[\Omega]+\Gamma_{R}[\Omega]\bigr)\,G_{0}^{a}[\Omega]
\ee
then we can write
\be
\bar{I}_{L}=-\bar{I}_{C}-\bar{I}_{R}
\ee
which is a consequence of energy conservation and $\bar{I}_{C}=i\, \Omega \,{\rm Tr} \bigl[(G_{0}^{r}[\omega]-G_{0}^{a}[\Omega])f_{0}f_{0}^{\dagger} \bigr]$.

It is important to note that the above expression (Eq.~(29)) contains $G_{0}^{r}$, $\Gamma_{L}$ which are independent of temperature. So in the ballistic case the current is independent of the temperature of the heat bath. However, the higher moments of current, for example $\langle I^{2}_{L} \rangle$, in general do depend on temperature. 

For the Hamiltonian given in Eq.~(2) it is also possible to calculate work done by the external time-dependent force which is given by $W=-\int_{0}^{\tau} dt \dot{f}^{T}(t) u_{c}(t)$ where the dot refers to derivative with respect to time. Using this definition one can then calculate $\langle W \rangle$ and $\langle W^{2}\rangle$ and can verify JE \cite{fluc3}. Following this definition, $P(W)$ is Gaussian and equivalent statement of JE classically reduces to $\langle W \rangle=\frac{\beta}{2}\,\bigl[\langle W^{2}\rangle-\langle W \rangle^{2}\bigr]$. Since the integration is over a time period $\tau=2\pi/\Omega$ the initial and final equilibrium states are the same and hence $\Delta F=0$. It is also important to realize that if we define $W'=-\int_{0}^{\tau} I_L(t)\,dt $ then it does not satisfy Jarzynski equality and $P(W')$ is not Gaussian. However the relation between first and second moment come out to be the same classically. 

The first and second moment of $W'$ (only the driven force contribution) can be written down explicitly
\begin{eqnarray}
\langle W' \rangle &=& \tau \,\Omega \, S[\Omega] \nonumber \\
\langle W'^{2}\rangle - \langle W' \rangle^{2}&=& \tau \,\hbar \, \Omega^{2}\, S[\Omega] \,\Bigl[\bigl(1+2\,f_{L}(\Omega)\bigr)-2\,{\rm T}[{\Omega}] \times \nonumber \\
&&\bigl(f_{L}(\Omega)-f_{R}(\Omega)\bigr)\Bigr].
\end{eqnarray}
When the leads are at the same temperature, we have,
\begin{equation}
\langle W'^{2}\rangle - \langle W' \rangle^{2}=\hbar \, \Omega \bigl(1+2\, f_{L}(\Omega)\bigr)\,\langle W' \rangle,
\end{equation}
which classically reduces to $\langle W'^{2}\rangle - \langle W' \rangle^{2}=\frac{2}{\beta}\,\langle W' \rangle$.
\subsection{Application to 1D chain}
\subsubsection{Rubin bath}
Here we consider a 1D chain with inter-particle spring constant $K$. We divide the full infinite system into three parts, the center, the left and the right lead. The leads are at the same temperature with the center. We drive the center with the force $f(t)$ and evaluate Eq.~(\ref{final1}). The classical equation of motion for the center atoms is given by
\be
\ddot{u}_j=K \bigl (u_{j-1}-2 u_{j}+u_{j+1} \bigr )+f_{j}(t), ~~~~ 1 \le j \le N_{C} 
\ee 
where $N_{C}$ is the number of particles in the center. The leads obey similar equations with $f_{j}(t)=0$. The equilibrium Green's functions satisfy time-translational invariance and hence Fourier's transform exists. In frequency space the retarded Green's function for the semi-infinite linear chain can be obtained by solving \cite{wang07} $[(\Omega+i\eta)^{2}-\tilde{K}]G_{0}^{r}=I$, where matrix $\tilde{K}$ which is infinite in both directions is $2K$ on the diagonal and $-K$ on the first off-diagonals.  The solution is translationally invariant in space index and is given by 
\be
G_{0,jk}^{r}[\Omega]=\frac{\lambda^{|j-k|}}{K(\lambda-\frac{1}{\lambda})},
\ee 
with $\lambda=-\frac{\bar{\omega}}{2K}\pm \frac{1}{2K}\sqrt{\bar{\omega}^{2}-4K^{2}}$ and $\bar{\omega}=(\Omega+i\eta)^{2}-2K$, Choosing between plus and minus sign by $|\lambda|\le 1$. The surface Green's function in frequency space is given by $\Sigma_{L}^{r}[\Omega]=-K \lambda$. It is clear from the expression of $\lambda$ that it is complex within the range $0 \le \Omega \le 2\sqrt{K}$ and is real outside this range. Hence $\Gamma_{L}$ is zero outside the phonon band.

\begin{figure}
\includegraphics[width=\columnwidth]{fig3.eps}%
\caption{\label{fig3} Energy current $\bar{I}_{L}$ as a function applied frequency for different system size of one-dimensional chain with force $f_{j}(t)=(-1)^{j}f_{o}e^{-i\Omega t}+c.c$. (a) $N_{C}$=4, (b) $N_{C}$=6, (c) $N_{C}$=8, (d) $N_{C}$=10. K=1 eV/(u\AA$^2)$.}
\end{figure}

\begin{figure}
\includegraphics[width=\columnwidth]{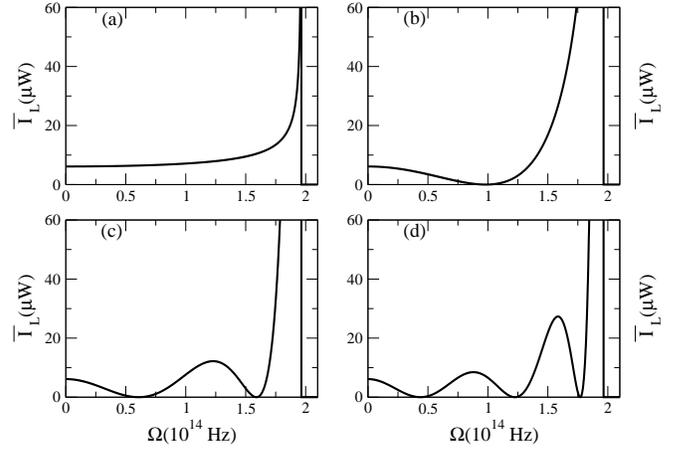}%
\caption{\label{fig4} Energy current $\bar{I}_{L}$ as a function applied frequency for different system sizes of one-dimensional linear chain with force $f_{j}(t)=(-1)^{j}f_{o}e^{-i\Omega t}+c.c$. (a) $N_{C}$=1, (b) $N_{C}$=3, (c) $N_{C}$=5, (d) $N_{C}$=7. K=1 eV/(u\AA$^2)$.}
\end{figure}

Here we consider the force $f_{j}(t)=f^{j}_{o}e^{-i\Omega t}+c.c.$ where $f^{j}_{o}=(-1)^{j}f_{0}$ which also mimic the structure of a crystal having alternate charges at the sites. For this force the expression for current is 
\begin{equation}
\bar{I}_{L}=
\begin{cases}
{-\frac{\Omega f_{o}^{2}}{2K}\, \frac{\bigl(1-(-1)^{N_C}\cos(N_{C}\,q)\bigr)}{\sin q \,\bigl(1+\cos q\bigr)}}, ~~~~~~~~ {\rm for}~ 0 \le \Omega \le 2\sqrt{K},   \\
{0}, ~~~~~~~~~~~~~~~~~~~~~~~~~~~~~~~~~~~{\rm for}~\Omega \ge 2 \sqrt{K}.
\end{cases}
\label{current_oned}
\end{equation}
 
\begin{figure}
\includegraphics[width=\columnwidth]{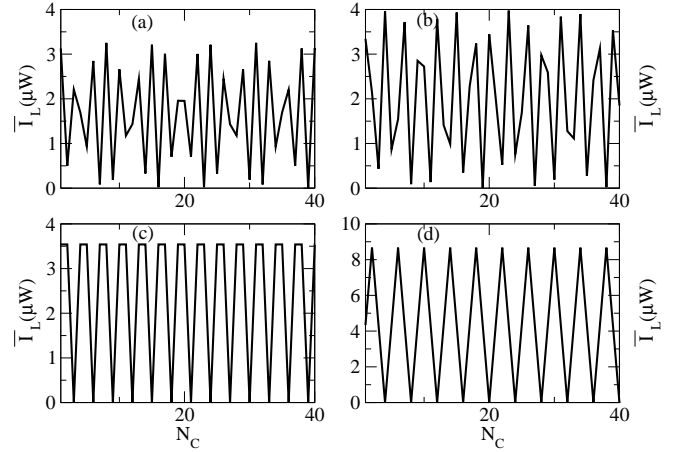}%
\caption{\label{fig5} Energy current $\bar{I}_{L}$ versus length of the center for different applied frequencies for one-dimensional linear chain. Here (a) $\Omega$=0.39, (b) $\Omega$=0.78, (c) $\Omega$=0.98, (d) $\Omega$=1.40. The frequencies are given in $10^{14}$(Hz) unit. The other parameters same as in Fig. 3 .}
\end{figure}
\noindent
$\bar{I}_{L}$ is of order 1 and $q$ is given by the dispersion relation $\Omega^{2}=2K(1-\cos q)$.\\
In Fig.~3 and Fig.~4, we plot energy current as a function of applied frequency for different system size. The value of force constant is chosen as $K=1$ eV/(u\AA$^2)$ and $f_{0}=1\,$nN in all our calculation. In Fig.~4, the current is nonzero even at zero frequency because the system as a whole is not charge neutral. For $N_{C}=1$ the current $\bar{I}_{L}$ is proportional to the density of states (DOS). More importantly the current is exactly zero when the applied frequency matches with the normal mode frequency of the system and the corresponding wave number is given by for even $N_{C}$, $q=2\pi n/N_{C}$ and for odd $N_{C}$, $q=(2n+1)\pi/N_{C}$ with $n=0,1,...,N_{C}-1$ . Therefore the number of resonance peaks and number of zero's depends on the eigenfrequencies and hence on the size of the center system. The average current diverges at $\Omega=2 \sqrt{K}$ as the DOS of the full system diverges at the maximum frequency of the whole system. For $\Omega \ge 2 \sqrt{K}$ the system does not allow energy to pass through.
\noindent
Similarly one can calculate the right lead current $I_{R}$ and the expression is the same with Eq.~(\ref{current_oned}). Since we apply force on all the atoms of the center by symmetry argument we can say that the total input current $I_C$ divides into two equal parts and goes into the leads i.e. $|I_{L}|=|I_{R}|=|I_{C}|/2$.

In Fig.~5, we give results for energy current as a function of total number of particles in the center for different values of external frequency. For finite systems the current oscillates with system size and depending on the values of $\Omega$ it shows periodicity with respect to $N_{C}$. The maximum amplitude of the average current is fixed and is proportional to $\Omega f_{o}^{2}/2 K$.

\subsubsection{Ohmic bath}
Here we consider the center system to be connected with two Ohmic baths. The difference between Rubin and Ohmic bath is that, the self energy in this case is approximated as $\Sigma[\Omega]=i \gamma \Omega$ where $\gamma$ is the friction coefficient. More precisely the $\Sigma_{L}$ and $\Sigma_{R}$ matrices are given by
\begin{eqnarray}
\Sigma_{L}^{lm}&=&i \, \gamma \, \Omega \,\delta_{lm} \, \delta_{l1}, \nonumber \\
\Sigma_{R}^{lm}&=&i \, \gamma \, \Omega \, \delta_{lm} \, \delta_{lN}.
\end{eqnarray}
\begin{figure}
\includegraphics[width=\columnwidth]{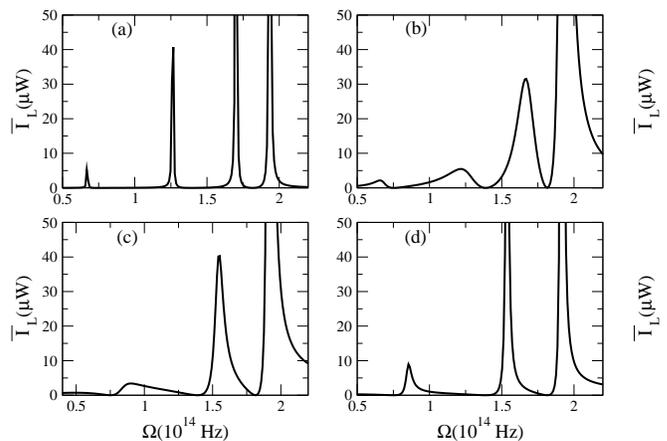}%
\caption{\label{fig3} Energy current $\bar{I}_{L}$ as a function applied frequency for different values of friction coefficient $\gamma$ of one-dimensional linear chain with force $f_{j}(t)=(-1)^{j}f_{o}e^{-i\Omega t}+c.c$. (a) $\gamma$=0.01, (b) $\gamma$=0.5, (c) $\gamma$=3.0, (d) $\gamma$=5.0. K=1 eV/(u\AA$^2)$ and $N_{C}=$8.}
\end{figure}
\noindent
Using the form of the Green's function in Eq.~(27) and after some bit of algebraic simplifications, we obtain the following results
\be
\bar{I}_{L}=2 \,\gamma\, \Omega^{2}\, f_{0}^{2}\, |g[\Omega]|^{2}
\ee
where 
\be
g[\Omega]=\sum_{j=1}^{N_{C}} (-1)^{j+1}G^{r}_{1j}[\Omega]
\ee
From the above expression and Eq.~(27) it is clear that energy current depends on the denominator $ A[\Omega]=|{\rm det}\bigl[D[\Omega]\bigr]|^{2}$ where $D[\Omega]=(\Omega^{2}\,I-K^{C}-i\Omega\,\Gamma_{L}-i\Omega\,\Gamma_{R})$ is $N_{C}\times N_{C}$ matrix. The matrix elements are given by $D_{ij}=\delta_{i,j}\bigl(\Omega^{2}-2\,K-i\Omega\,\gamma(\delta_{i,1}+\delta_{i,N})\bigr)-K\,\delta_{i,j+1}-K\,\delta_{i,j-1}$. If we denote $P_{N_C}[\Omega]={\rm det}(\Omega^{2}-K^{C})$ to be the characteristic polynomial of the matrix $K^{C}$ with $N_{C}$ particles then it can be shown that \cite{ren2}
\be
A[\Omega]=\bigl[P_{N_C}[\Omega]-\gamma^{2} \,\Omega^{2} P_{N_{C}-2}[\Omega]\bigr]^{2}+ 4 \gamma^{2}\,\Omega^{2}\,P_{N_{C}-1}^{2}[\Omega],
\ee
where $P_{N_{C}-1}[\Omega]$ is the polynomial of the $(N_{C}-1)\times (N_{C}-1)$ force constant matrix $K^{C}$ with first row and column or last row and column taken out from $K^{C}$ and similarly $P_{N_{C}-2}[\Omega]$ is the polynomial of the $(N_{C}-2)\times (N_{C}-2)$  matrix by taking out the first and last rows and columns from $K^{C}$. The resonance and the zero's of current corresponds to the minimum and maximum value of $A[\Omega]$ respectively. It is difficult to obtain explicit solution in this case. However the equation become simple for small and large value of $\gamma$, the friction coefficient. For small friction it is clear from Eq.~(40) that $A[\Omega]=P_{N_{C}}^{2}[\Omega]$. So the resonant frequencies depends on $N_{C}$ eigenfrequencies of the force constant matrix $K^{C}$. In the opposite limit i.e, for large $\gamma$ we obtain $A[\Omega]=P_{N_{C}-2}^{2}[\Omega]$. So depending on the value of $\gamma$ the resonance peaks shift from $N_{C}$ to $N_{C}-2$. 

In Fig.~6, we plot the current with applied frequency for different values of damping coefficient $\gamma$. The value of $\gamma$ is chosen in proper units. The zero values of the current is same as in Rubin's case. However there is a gradual shift in the resonance peak depending on the parameter $\gamma$. The current doesn't diverge at $\Omega=2 \sqrt{K}$ and the width of the peaks depends of $\gamma$. We check numerically the behavior of $\bar{I}_{L}$ with system length and we found that the behavior is similar with Rubin baths. In this case also we have $|I_{L}|=|I_{R}|=|I_{C}|/2$.

Similar Ohmic model was also investigated by Marathe et.~al \cite{Marathe} for $N_{C}=2$ where they conclude that this model cannot work either as a heat pump or as a heat engine. Our calculation agrees with their results. 

It is also possible to calculate current in the overdamped regime by dropping the term $(\Omega+ i\eta)^{2}$ in $G^{r}[\Omega]$ given in Eq.~(27). In this regime for $N=1$ our result agrees with the result obtained in Ref.~\onlinecite{fluc4} for magnetic field $B=0$. 

\subsubsection{Comparison between Rubin and Ohmic bath for driving force on single site}
As we have seen that if we apply force on all the atoms of the center because of the symmetry of the problem if we interchange the left and right lead (which we assume to be the same) the value of the current should not change and hence we have the only possible solution $|I_{L}|=|I_{R}|=|I_{C}|/2$. But this is not the case, at least for Ohmic bath, if we apply force on a single or multi-particles but not on all. 

If we consider the force on the $\alpha$th particle as $f_{i}(t)=\delta_{i\alpha}\bigl(f_{0}^{i}\, e^{-i \Omega t}+ c.c \bigr)$ then for the Rubin bath case using Eq.~(28) and Eq.~(29) we get 
\begin{eqnarray}
\bar{I}_{L}&=&-2 \, \Omega \,K \,{\rm Im}(\lambda)\, f_{0}^{\alpha}\,(f_{0}^{\alpha})^{*} \, |G_{0,\alpha 1}|^{2} \nonumber \\
\bar{I}_{R}&=&-2 \, \Omega \,K \,{\rm Im}(\lambda)\, f_{0}^{\alpha}\,(f_{0}^{\alpha})^{*} \, |G_{0,N \alpha}|^{2}
\end{eqnarray}
Using the solution for $G_{0}^{r}[\Omega]$ given in Eq.~(35) we obtain
\be
\bar{I}_{L}= \bar{I}_{R}=
\begin{cases}
{-\frac{\Omega}{2K \sin q} f_{0}^{\alpha}\,(f_{0}^{\alpha})^{*}},~~~~~~~~ {\rm for}~ 0 \le \Omega \le 2\sqrt{K},   \\
{0}, ~~~~~~~~~~~~~~~~~~~~~~~~~{\rm for}~\Omega \ge 2 \sqrt{K}.
\end{cases}
\ee 
which says that, because the full system is translationally invariant in space, the magnitude of current does not depend on which site the force is applied and hence $|I_{L}|=|I_{R}|=|I_{C}|/2$ is the only possible solution. The result is similar with $N_{C}=1$ in Eq.~(\ref{current_oned}). 

However,this scenario is not valid for Ohmic bath. In this case the full translational symmetry is broken and hence applying force on different sites generate different magnitudes of current on left and right lead. In Fig.~7, we plot the heat current $\bar{I}_{L}$ and  $\bar{I}_{R}$ for one-dimensional chain as a function applied driving frequency at different sites. Clearly $\bar{I}_{L}$ and  $\bar{I}_{R}$ are different in magnitudes. Hence by applying force on different sites it is possible to control current in both the leads for Ohmic case.
\begin{figure}
\includegraphics[width=\columnwidth]{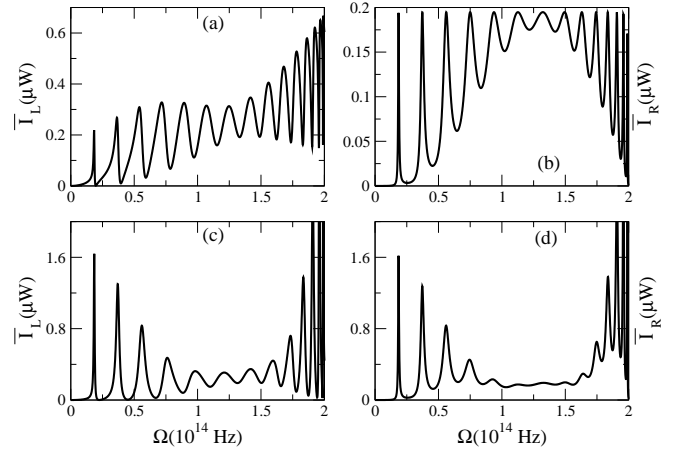}%
\caption{\label{fig3} Energy current $\bar{I}_{L}$ and  $\bar{I}_{R}$ as a function of applied frequency for driven force at different site of one-dimensional chain connected to Ohmic bath. (a) and (b) are for $\alpha$=1 and (c) and (d) are for $\alpha=3$, $N_{C}=$16. $K=1$ eV/(u\AA$^2)$.}
\end{figure}

\subsubsection{Heat pump}
Heat pump by definition transfers heat from cooler region to hotter region. One-dimensional linear system with force applying on any number of sites fails to work as a heat pump. To understand the reasoning we may consider Eq.~(23) which says that heat current $\bar{I}_{L}$ is a sum of two parts. If we assume $T_L>T_R$ then the first term in Eq.~(23) which gives the steady state heat flux due to temperature difference is positive, i.e, current goes from left to right lead and the driving term which does not depend on temperature, always contribute a negative value to both $\bar{I}_{L}$ and $\bar{I}_{R}$. Hence $\bar{I}_{R}$ is always negative independent of whether we apply force on one site or on all the sites. So it is not possible to transfer heat from right lead to left lead in this case.

\subsection{Application to 2D square lattice : Rubin Bath}
In this case we consider a square lattice with force constant $K$ both in $x$ and $y$ direction. We take a small part of the full infinite system which is square in shape and call it the center and rest is treated as a bath and is kept at a constant temperature with the center. The classical equation of motion for the $x$-component of the center atoms is given by
\begin{eqnarray}
\ddot{u}^{x}_{j,k}&=&Ku^{x}_{j-1,k}-4Ku^{x}_{j,k}+Ku^{x}_{j+1,k}+Ku^{x}_{j,k+1} \nonumber \\
+&&Ku^{x}_{j,k-1}+f^{x}_{k}(t),  ~~~~~~~~~~~~~~ 1 \le j,k \le N_{C};  \nonumber \\
\end{eqnarray}
and similar equation for the $y$-component. The total number of particles in the center is $N_{C}^{2}$. The retarded Green's function for the full system is given by \cite{Maradudin}
\be
G^{r}_{{\it {l}}{\it {l}'}}[\Omega]= \frac{1}{N^{2}}\sum_{\bf k} \frac{e^{i ({\bf R}_{l}-{\bf R}_{l'}). {\bf k}}}{\Omega^{2}-2K\,[2-\cos(k_{x}a)-\cos(k_{y}a)]},
\ee
where $l=l_{2}+(l_{1}-1)N$ and $l'=l_{2}'+(l_{1}'-1)N$, ${\bf{R}}_{l}=l_{1}{\bf{a}}_{1}+l_{2}\bf{a}_{2}$, ${\bf{R}}_{l'}=l'_{1}{\bf{a}}_{1}+l'_{2}{\bf{a}}_{2}$ and $N$ is the total number of particle in the full system, ${\bf k}$ is the wave-vector and it's components are given by $k_{x}=\frac{2 \pi n_{x}}{N a}$, $k_{y}=\frac{2 \pi n_{y}}{N a}$ where $a$ is the lattice constant. ${\bf{a}}_{1}$ and ${\bf{a}}_{2}$ are the primitive lattice vectors.
In the large $N$ limit, i.e., $N \rightarrow \infty$, one can write 
\be
G^{r}_{{\it {l}}{\it {l}'}}[\Omega]=\frac{1}{2\pi}\int_{-\pi}^{\pi} dq_{y}\, \cos(n_{2}q_{y})\, L(q_{y}),
\ee
where  $n_{1}=l'_{1}-l_{1}$, $n_{2}=l'_{2}-l_{2}$ and 
\be
L(q_{y})=\frac{1}{2\pi}\int_{-\pi}^{\pi} dq_{x} \, \frac{e^{i n_{1}q_x}}{\Omega^{2}-2K\,\bigl[2-\cos(q_{x})-\cos(q_{y})\bigr]}.
\ee
This integral can be done using the contour integration technique and can be written as
\begin{equation}
L(q_{y})=\frac{\lambda^{n_{1}}}{K (\lambda-\frac{1}{\lambda})},
\end{equation}
with $\lambda=-\frac{\bar{\omega}}{2 K}\pm \frac{\sqrt{(\bar{\omega}^{2}-4 K^{2})}}{2 K}$ and $\bar{\omega}=\Omega^{2}-2 K\,\bigl(2-\cos(q_{y})\bigr)$. The choice between plus and minus sign depends on $|\lambda| \le 1$. In this case the explicit expression for $\Sigma_{L}^{r,a}[\Omega]$ is not known. However, by knowing $G^{r}_{{\it {l}}{\it {l}'}}[\Omega]$ we can compute the self energy of the infinite 2D square lattice with a removed square part using the following equation
\begin{equation}
\Sigma_{L}^{r}[\Omega]=(\Omega+i\eta)^{2}-K_{C}-\bigl[G_{0}^{r}[\Omega]\bigr]^{-1}.
\end{equation}
We can obtain $\Gamma_{L}$ from the above expression as $\Gamma_{L}[\Omega]=-2\, {\rm{Im}}\bigl(\Sigma_{L}^{r}[\Omega]\bigr)=2 \,{\rm{Im}}\bigl [G_{0}^{r}[\Omega]\bigr]^{-1}$.

\begin{figure}
\includegraphics[width=\columnwidth]{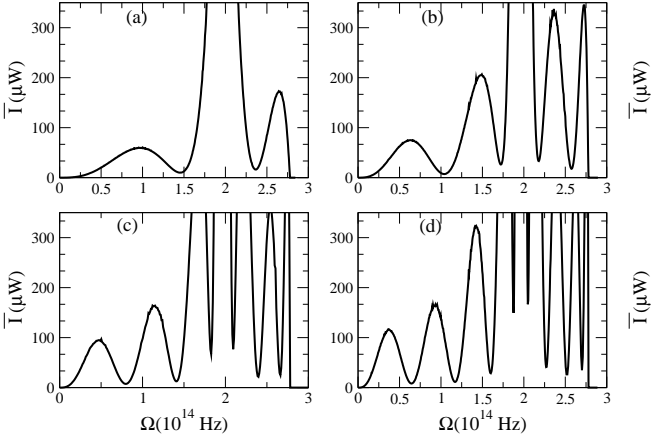}%
\caption{\label{fig4} Energy current $\bar{I}$ as a function applied frequency for different system size of square lattice with force ${\bf{F_{j}(t)}}={\bf (-1)^{j}f_{0}}e^{-i\Omega t}+c.c$. (a) $N_{C}$=4, (b) $N_{C}$=6, (c) $N_{C}$=8, (d) $N_{C}$=10. $K=1$ eV/(u\AA$^2)$.}
\end{figure}

In Fig.~8 we plot the average current going out of the center with frequency where the force is in both in $x$ and $y$ direction with same magnitude ($f_{0}=1\,$nN) and is given by ${\bf{f_{j}(t)}}={\bf (-1)^{j} f_{0}}e^{-i\Omega t}+c.c$. The behavior of average current in this case is quite similar to the 1D case. The oscillation also increases with $N_{C}$ and the value of current goes to minimum when the applied frequency matches with the normal mode frequencies of the full system. 

In Fig.~9 we plot the current with system size and it is found that the current oscillates with $N_{C}$ and it also shows a periodic pattern depending on the value of $\Omega$. $\bar{I_L}$ is roughly proportional to $N^{2}_{C}$, the total number of particles in the center.

\begin{figure}
\includegraphics[width=\columnwidth]{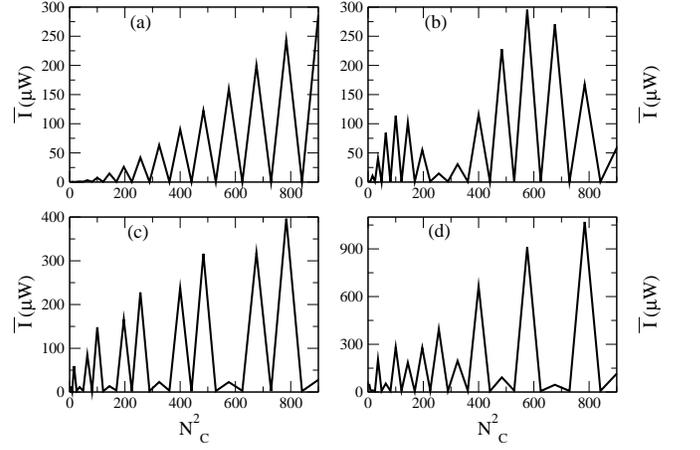}%
\caption{\label{fig5} Energy current $\bar{I}$ versus size of the center for different applied frequencies for two-dimensional square lattice. Here (a) $\Omega$=0.10, (b) $\Omega$=1.47, (c) $\Omega$=0.98, (d) $\Omega$=0.39. The frequencies are given in units of $10^{14}$(Hz). The other parameters same as in Fig.~3.}
\end{figure}
\section{Nonlinear interaction at the center}
 
It is possible to study the effect due to nonlinear interaction in the center for this model. In this case we assume that both the force and the cubic interaction switched on at $t=-\infty$. So using contour-ordered Green's function and interaction picture the center Green's function can be written as
\begin{eqnarray}
G_{jl}(\tau,\tau')&=&-\frac{i}{\hbar}\langle T_{\tau} u_{j}(\tau) u_{l}(\tau') e^{\sum_{m}\frac{i}{\hbar} \int d\tau'' f_{m}(\tau'') u_{m}(\tau'')}\nonumber\\
&&e^{-\frac{i}{\hbar} \int d\tau'''H_{n}^{I}(\tau''')} \rangle_{G_{0}}.
\end{eqnarray}
We take cubic potential which is 
\be 
H_{n}(\tau)=\frac{1}{3}\sum_{jkl}\int d\tau' \int d\tau'' T_{jkl}(\tau,\tau',\tau'') u_{j}(\tau)u_{k}(\tau')u_{l}(\tau'')
\ee
where $T_{jkl}(\tau,\tau',\tau'')=T_{jkl}\delta(\tau,\tau')\delta(\tau,\tau'')$. If we expand the nonlinear interaction the first term gives us our old linear result. The first nonzero contribution comes from the second term of both the nonlinear potential and the force. The expression for the Green's function in first order of force is given by
\begin{eqnarray}
&&G_{mn}(\tau,\tau')=-\frac{1}{3}(-\frac{i}{\hbar})^{3}\int d\tau''\int d\tau_{1} \int d\tau_{2} \int d\tau_{3} \nonumber \\
&& \sum_{jklo}\langle T_{\tau} u_{m}(\tau) u_{n}(\tau') f_{o}(\tau'') u_{o}(\tau'')\nonumber \\
&&T_{jkl}(\tau_{1},\tau_{2},\tau_{3}) u_{j}(\tau_{1})u_{k}(\tau_{2})u_{l}(\tau_{3})\rangle_{G_{0}}.
\end{eqnarray}
Since the density operator is quadratic we can use Wick's theorem and finally we get 15 possible terms which gives rise to three independent Feynman diagrams. The final expression combining all these diagrams is
\begin{eqnarray}
&&I_{L}(t)=\frac{1}{4 \pi^{2}}\sum_{jklmno} T_{jkl}\int_{-\infty}^{\infty} d\omega' \int_{-\infty}^{\infty} d\omega''\hbar \omega' e^{-i \omega''t}G_{jo}^{r}[\omega'']\nonumber \\
&&f_{o}[\omega'']\Big(G_{ml}^{r}[\omega'+\omega'']G_{kn}^{r}[\omega']\bar{\Sigma}_{nm}[\omega']+G_{ml}^{r}[\omega'+\omega'']\nonumber \\
&&\bar{G}_{kn}[\omega']\Sigma_{nm}^{a}[\omega']+\bar{G}_{ml}[\omega'+\omega'']G_{kn}^{a}[\omega']\Sigma_{nm}^{a}[\omega']\Big)
\end{eqnarray}
From this expression it is clear that in the steady state there is no contribution to current  to the linear order in $f$ if we consider a cubic inter atomic potential. To see the effect due to nonlinearity and also temperature dependent heat current it is important to go to higher order in force and also of the nonlinear potential. 

\section{conclusion}
In summary, we present an exact analytical expression of energy current for driven linear system in time domain. The energy current is written in terms of the displacement of the center atoms and self energy of the heat bath. We study the properties of energy current for two different types of heat baths with different forms of self energy $\Sigma$. We obtain an explicit expression of current for one-dimensional linear chain, connected to Rubin baths, exploring the translational symmetry of the full system. We discuss the similarities and differences between Rubin and Ohmic bath when the force is applied on all sites or on single site. We also relate the time integral of left lead current with work and discuss that this particular definition does not obey JE. However, we find that the relation between first and second moment of work in both cases are same, classically. It will be interesting to study the general features of current using Eq.~(25) with other forms of time dependent forces. The effect on current and heat pumping due to nonlinear interaction in higher order of force are worthy of further explorations.  
 
\section*{ACKNOWLEDGEMENTS}
We are grateful to Jin-Wu Jiang, Juzar Thingna, Meng Lee Leek and Eduardo C. Cuansing for insightful discussions. This work is supported in part by URC grant R-144-100-257-112 of National University of Singapore.


\begin{thebibliography}{99}
\bibitem{casati} M. Terraneo, M. Peyrard, and G. Casati, Phys. Rev. Lett. {\bf 88}, 094302 (2002);
\bibitem{baowen1}B. Li, L. Wang, and G. Casati, Phys. Rev. Lett. {\bf 93}, 184301 (2004).
\bibitem{baowen2} B. Li, L. Wang, and G. Casati, Appl. Phys. Lett. {\bf 88}, 143501 (2006).
\bibitem{chang} C. W. Chang, D. Okawa, H. Garcia, A. Majumdar, and A. Zettl, Phys. Rev. Lett. {\bf 99}, 045901 (2007).
\bibitem{xie} R.-G. Xie, C.-T. Bui, B. Varghese, M.-G. Xia, Q.-X. Zhang, C.-H. Sow, B. Li, and J. T. L. Thong, Adv. Funct. Mat {\bf 21}, 1602 (2011).
\bibitem{wang06}J.-S. Wang, J. Wang and N. Zeng, Phys. Rev. B \textbf{74}, 033408 (2006). 
\bibitem{wang08}J.-S. Wang, J. Wang, and J. T. L\"u, Eur. Phys. J. B, \textbf{62}, 381 (2008). 
\bibitem{eduardo}E. C. Cuansing and J.-S. Wang, Phys. Rev. E \textbf{82}, 021116 (2010). 
\bibitem{eduardo1}E. C. Cuansing and J.-S. Wang, Phys. Rev. B \textbf{81}, 052302 (2010).
\bibitem{Dhar}A. Dhar, Adv. in Phys., \textbf{57}, 457-537 (2008).
\bibitem{Dhar1}A. Dhar and D. Roy, J. Stat. Phys, \textbf{125}, 4, (2006).
\bibitem{Tanimura}Y. Tanimura,  J. Phys. Soc. Jpn. \textbf{75},  082001, (2006).
\bibitem{gaspard}M. Esposito and P. Gaspard, Phys. Rev. E, \textbf{76}, 041134, (2007).
\bibitem{Lepri}S. Lepri, R. Livi, and A. Politi, Phys. Rep. \textbf{377}, 1 (2003).
\bibitem{Lebowitz} F. Bonetto, J. L. Lebowitz, and L. Rey-Bellet, ``Fourier's Law: A Challenge to Theorists,'' {\it {Mathematical Physics 2000}} (Imp. Coll. Press, London, 2000)
\bibitem{Ren} J. Ren and B. Li, Phys. Rev. E {\bf 81}, 021111 (2010).
\bibitem{Li1} N. Li, P. H\"anggi, and B. Li, Europhys.~Lett. \textbf{84}, 40009
(2008).
\bibitem{Li2} N. Li, F. Zhan, P. H\"anggi, and B. Li, Phys. Rev. E {\bf
80}, 011125 (2009).

\bibitem{saito} K. Saito, EPL, \textbf {83}, 50006 (2008).
\bibitem{Harada}T. Harada, and S. I. Sasa, Phys. Rev. E \textbf{73}, 026131 (2006).
\bibitem{JE} C. Jarzynski, Phys. Rev. Lett. \textbf{78}, 2690 (1997).
\bibitem{JE1}C. Jarzynski, Phys. Rev. E \textbf{56}, 5018 (1997).
\bibitem{fluc2}A. Dhar, Phys. Rev. E \textbf{71}, 036126 (2005).
\bibitem{fluc3}T. Mai and A. Dhar, Phys. Rev. E \textbf{75}, 061101 (2007). 
\bibitem{fluc4}A. M. Jayannavar and M. Sahoo, Phys. Rev. E \textbf{75}, 032102 (2007).
\bibitem{Rubin}R. J. Rubin and W. L. Greer, J. Math. Phys. {\bf 12}, 1686 (1971). 
\bibitem{Weiss}U. Weiss, {\it{Quantum Dissipative Systems}}, 2nd edn. (World Scientific, 1999).
\bibitem{schwinger-keldysh} J. Schwinger, J. Math. Phys. {\bf 2}, 407
  (1961); L.V. Keldysh, Sov. Phys. JETP {\bf 20}, 1018 (1965).
\bibitem{rammer86} See, for a review, J. Rammer and H. Smith, Rev. Mod.
  Phys. {\bf 58}, 323 (1986).
\bibitem{jauho94} A.-P. Jauho, N.S. Wingreen, and Y. Meir, Phys. Rev.
  B {\bf 50}, 5528 (1994).
\bibitem{wang07}J.-S. Wang, N. Zeng, J. Wang, and C. K. Gan, Phys. Rev. E {\bf 75}, 061128 (2007). 
\bibitem{ren2} S. Zhang, J. Ren, and B. Li, arxiv:1102.4113
\bibitem{Marathe}R. Marathe, A. M. Jayannavar, and A. Dhar, Phys. Rev. E {\bf 75}, 030103(R) (2007).
\bibitem{Maradudin}A. A. Maradudin, P. Mazur, E. W. Montroll, and G. H. Weiss, Rev. Mod. Phys. {\bf 30}, 175–196 (1958).  

\end{thebibliography}
\end{document}